\documentclass[reprint,prl,amsmath,amssymb,floatfix,nolongbibliography]{revtex4-2}

\usepackage{color}
\usepackage{multirow}
\usepackage{bm}
\usepackage{epsfig}
\usepackage{psfrag}
\usepackage[latin1]{inputenc}
\usepackage[english]{babel}
\usepackage{amsfonts}
\usepackage{amsmath}
\usepackage{upgreek}
\usepackage[hidelinks]{hyperref}
\usepackage[dvipsnames]{xcolor}
\usepackage[normalem]{ulem}
\usepackage{bbold}
 
\usepackage{pdfpages} 
\makeatletter
\AtBeginDocument{\let\LS@rot\@undefined}
\makeatother

\DeclareSymbolFont{bbold}{U}{bbold}{m}{n}
\DeclareSymbolFontAlphabet{\mathbbold}{bbold}

\newcommand{\e}{{\rm e}}

\newcommand{\ex}[1]{\langle #1 \rangle}

\begin{document}

\title{
	Finite Soliton Width Matters: Investigating Non-equilibrium  Exchange Phases of Anyons
	}
 
\author{
  Matthias Thamm  
	}  
\author{
	Bernd Rosenow
	}  
\affiliation{
	Institut f\"{u}r Theoretische Physik, Universit\"{a}t
  Leipzig,  Br\"{u}derstra{\ss}e 16, 04103 Leipzig, Germany
	} 
	
\date{\today}

\begin{abstract}

Unlike bosons and fermions, quasi-particles in two-dimensional quantum systems, known as anyons, exhibit statistical exchange phases that range between $0$ and $\pi$. In fractional quantum Hall states, these anyons, possessing a fraction of the electron charge, traverse along chiral edge channels. This movement facilitates the creation of anyon colliders, where coupling different edge channels through a quantum point contact enables the observation of two-particle interference effects. Such configurations are instrumental in deducing the anyonic exchange phase via current cross-correlations. Prior theoretical models represented dilute anyon beams as discrete steps in the boson fields. However, our study reveals that incorporating the finite width of the soliton shape is crucial for accurately interpreting recent experiments, especially for collider experiments involving anyons with exchange phases $\theta>\pi/2$, where prior theories fall short.

\end{abstract}
 
\maketitle

\emph{Introduction---}Anyons are exotic quasi-particles of two dimensional systems that exhibit exchange statistics intermediate between bosons and fermions \cite{Leinaas.1977,Laughlin.1983,Halperin.1984,Arovas.1984,Feldman.2021}. They appear in the fractional quantum Hall (FQH) effect, where their charge $e^*$ is a fraction of the electron charge and exchanging two anyons contributes an exchange phase $\theta$.  While the fractional charge has been measured some time ago \cite{Reznikov.1999,Saminadayar.1997}, evidence for the anyonic statistical phase was only found recently \cite{Bartolomei.2020,Nakamura.2020,Nakamura.2022,Nakamura.2023,Lee.2023,Ruelle.2023,Glidic.2023}. The experiments \cite{Bartolomei.2020,Lee.2023,Ruelle.2023,Glidic.2023} are based on the idea that the signature of  anyonic statistics is imprinted in current cross-correlations in setups involving multiple quantum point contacts (QPCs)  \cite{Safi.2001,Vishveshwara.2003,Martin.2005,Kim.2005,Vishveshwara.2010,Campagnano.2012,Campagnano.2013,Rosenow.2016,Kim.2006,Han.2016,Lee.2019,Lee.2020}. The specific signatures predicted in \cite{Rosenow.2016} can be interpreted in terms of time domain interference of anyons  \cite{Han.2016,Lee.2019,Lee.2020,Schiller.2022}.

\begin{figure}[t!]
    \centering
    \includegraphics[width=8.6cm]{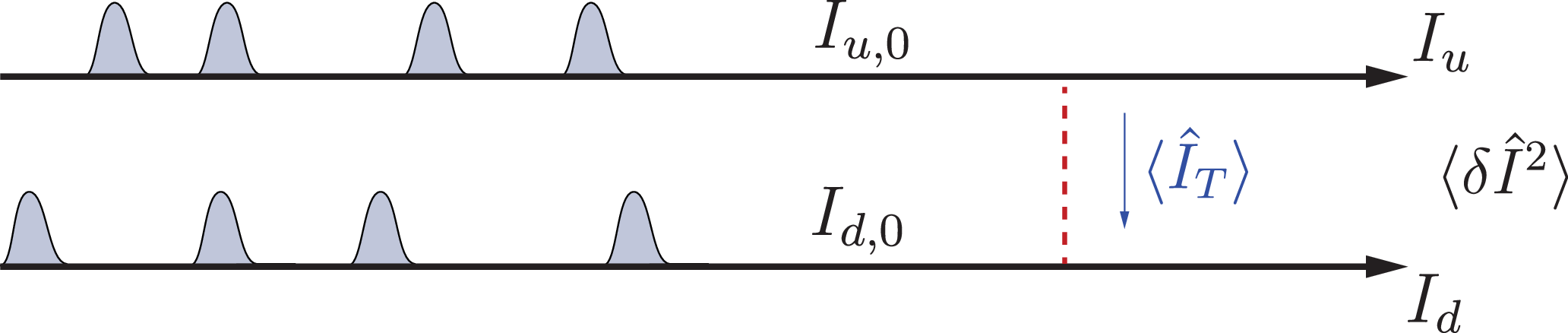}
    \caption{\label{Fig:sys}Schematic of an anyon collider setup with two chiral edge channels (black arrows) and a quantum point contact (red). We consider dilute beams of anyons  incident along the  edges (blue) which are allowed to tunnel between the edges at the quantum point contact with small tunneling amplitude. Treating the tunneling coupling as a small perturbation, we compute the tunneling current expectation value $\ex{\hat{I}_T}$ and the current cross correlations $\ex{\delta\hat{I}^2}$. We focus on the symmetric case where the current before the quantum point contact is the same on both edges, i.e., $I_-=I_{u,0}-I_{d,0}=0$.
 }
\end{figure}

An  observable  studied in  collision experiments \cite{Bartolomei.2020,Ruelle.2023,Glidic.2023} (for a schematic setup see Fig.~\ref{Fig:sys}) is the generalized Fano factor $P$, defined as the sum of the current noise due to partitioning at the collision QPC  and the noise in the incoming currents transmitted through the collision  QPC, normalized by the latter \cite{Rosenow.2016}. For a fractional quantum Hall edge with equilibrium density fluctuations, the correlation function of an anyon tunneling operator decays like a power law in time,  governed by a dynamical exponent $2\delta$. The Fano factor in a symmetric collider setup with two balanced incident anyon beams has been found to be $P(0)  = 1 - \tan(\theta) \tan(\pi\delta)^{-1} (1-2\delta)^{-1}$, where  a negative $P$ is a robust signature of anyonic statistics \cite{Rosenow.2016}. This form of the Fano factor applies to cases where $\delta<1/2$ and  $\theta<\pi/2$, with the restriction due to  the approximation of quasi-particle pulses with vanishing spatial and temporal width made in \cite{Rosenow.2016}. 

However, experiments indicate that the dynamical exponents $\delta$ can deviate from the universal value describing a pristine edge, and may be close to one \cite{Lee.2023,Ruelle.2023} even for filling  $\nu=1/3$ . As $\delta$ appears in the expression for the Fano factor, from which one would like to obtain the exchange phase $\theta$, an uncertainty in $\delta$ leads to an uncertainty in $\theta$. To reliably determine the exchange phase from experiments, it is therefore necessary to know the Fano factor as a function of  $\delta$ outside the range $\delta<1/2$.
In addition,  previous theory does not apply to the $\nu=2/5$ FQH state,  where $\theta=3\pi/5$ is expected for $e^* = e/5$ quasi-particles. The reason is that 
in the exponential  
$e^{i 2 \theta}$ a positive phase $2 \theta > \pi$ has a the same effect as a negative phase $2 \pi - 2 \theta$, giving rise to a positive $P(0)$ due to this ambiguity.  Experimentally, 
small negative  Fano factors have been measured \cite{Ruelle.2023,Glidic.2023}, and explaining this finding is a challenge for theory.

\begin{figure}[t!]
    \centering
    \includegraphics[width=5.3cm]{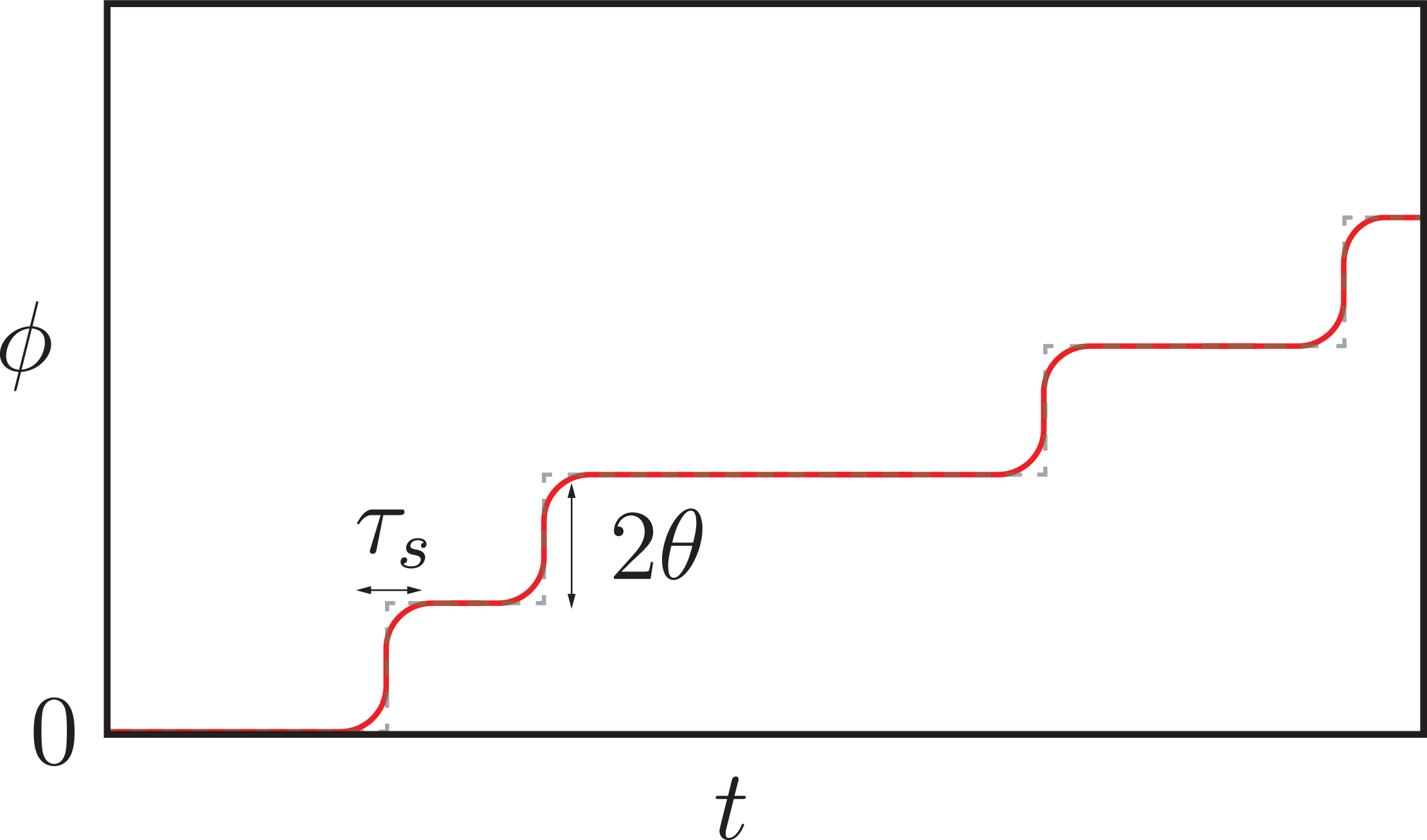}
    \caption{\label{Fig:BroadenedField}A step of height $2\theta$ in the boson field $\phi $ describes the passage of a quasi-particle with exchange phase $\theta$. Here, the anyon creation operator on the edge is related to the boson field via $\psi(x,t) \propto \e^{i \phi(x,t)}$. Instead of discrete steps (gray, dashed), we consider a smooth increase with soliton width $\tau_s$.
 }
\end{figure}

In this letter, we compute the Fano factor for dynamical exponents $0<\delta\leq 1$ for 
exchange phases $\theta = 3 \pi/5$ and $\theta = \pi/3$ by taking into account a 
finite width of anyonic pulses in the incident  dilute beams. This idea is based on the observation of Schiller et al.~\cite{Schiller.2022}
that a finite  width of quasi-particles   allows to correctly compute the Fano factor even for electrons with  $\theta = \pi $ and $\delta = 1$.  When computing the Fano factor, the time integral describing current cross correlations contains a factor $\sim (it+\tau_c)^{-2\delta}$ reflecting the equilibrium decay of the correlation function of 
the anyon tunneling operator. Hence, the integral  is dominated by large times for $\delta<1/2$, such that approximating the anyon pulse as zero width is a good approximation. In contrast, for  $\delta>1/2$ there are relevant contributions from short times where details on the shape of the solitons matter (Fig.~\ref{Fig:BroadenedField}). The resulting Fano factor for large dynamical exponents is non-universal depending on the width of anyon pulses.

\emph{Setup---}We consider an anyon collider  consisting of chiral edge channels ($\alpha\in\{u,d\}$) as schematically illustrated in Fig.~\ref{Fig:sys}, which can be realized in FQH systems. Here, the  dilute anyon beams impinging on the collision QPC  are created by  applying a voltage $V$ across  additional 
source QPCs (not shown in Fig.~\ref{Fig:sys}), allowing tunneling of quasi-particles into the $u,d$ edges  with a small tunneling probability $T_s\ll1$, thereby ensuring that the spatial separation between anyons is much larger than their width.  The edge states \cite{Halperin.1982} are described as chiral Luttinger liquids \cite{Wen.1990}.
Due to the presence of the collision QPC  between the $u$ and $d$ edge,  
time-domain interference effects allow the extraction of information about the anyon braiding phase from the tunneling current expectation value $\ex{\hat{I}_T}$ and current cross correlations $\ex{\delta \hat{I}^2}$.

In the bosonization formalism, the  operator for the tunneling current between  upper to the lower edge is given by
$ \hat{I}_T = i e^* (A^\dagger - A)$, with $
    A(t)= \zeta \e^{i\phi_u(0,t) - i\phi_d(0,t)}$, 
where $e^*$ is the fractional charge of the anyons, $\zeta$ is the tunneling amplitude at the collision QPC, and $\phi_\alpha(x,t)$ is the boson field for the anyons on edge $\alpha$. The charge density on the edges is given by $\rho_\alpha(x,t) = \partial_x \phi_\alpha(x,t)/2 \pi$. 
The collision QPC is located at $x=0$. The equal time commutator
$[\phi_\alpha(x,t), \phi_\beta(y,t)] = ie^* \pi\delta_{\alpha\beta} {\rm sgn}(x-y)$ is proportional to the fractional charge measured in units of the electron charge $e$, while the equal position commutator $[\phi_\alpha(x,t_1), \phi_\beta(x,t_2)] = i \delta \pi\delta_{\alpha\beta} {\rm sgn}(t_1-t_2)$ contains the dynamical exponent $\delta$ that can be different from $e^*$ and also $\theta/\pi$ \cite{Rosenow.2002}.

To model the  non-equilibrium situation depicted in Fig.~\ref{Fig:sys}, we decompose the boson fields into $\phi^{(0)}_\alpha(0,t)$ describing the
equilibrium quantum fluctuations and a classical non-equilibrium part describing Possionian fluctuations in the quasi-particle number with expectation value $\ex{I_{\alpha,0}}/e^*$ \cite{Rosenow.2016}. An anyon with exchange phase  $\theta$  arriving at time $t_0$ at the collision QPC causes a shift of the non-equilibrium part of the boson field, described  by 
\begin{align}
    \phi_\alpha(0,t) &\to \phi^{(0)}_\alpha(0,t) - 2\theta\left[\frac{1}{\pi}\arctan\left(\frac{t-t_0}{\tau_s}\right)-\frac{1}{2}\right] \ , \label{Eq:BosFieldSingPart}
\end{align}
where we take into account the finite width $\tau_s$ of the anyon pulses  \cite{Schiller.2022} (see Fig.~\ref{Fig:BroadenedField}). This finite soliton width has previously been neglected by using discrete steps in the boson field, which corresponds to delta peaks in the density.

\emph{Current and noise in linear response---}The expectation value of the tunneling current is given by
$  \ex{\hat{I}_T}(t) = e^* \int_{-\infty}^\infty \ex{[A^\dagger(0),A(t)]} {\rm d}t $.
The noise in the tunneling current is obtained as
 \begin{align}
    \ex{(\delta\hat{I})^2}(t) &\equiv  \ex{\,\big[\hat{I}_T(t) - \ex{\hat{I}_T}(t)\big]\,\big[\hat{I}_T(0) - \ex{\hat{I}_T}(0)\big]\,} , 
\end{align}
with the $\omega=0$ frequency component of its Fourier transform given by
$    \ex{(\delta\hat{I})^2}_{\omega=0} =  (e^*)^2 \int_{-\infty}^\infty \ex{\{A^\dagger(t) , A(0)\}}_0  {\rm d}t $.

Under the assumption that the beams are sufficiently dilute, i.e.\@ $\hbar/e^*V\ll e^*/I_{\alpha,0}$, we can treat the individual anyons on one edge as independent, allowing to describe them using Poisson statistics. This gives rise to the correlation function  
\begin{align}
    \frac{\ex{A^\dagger(0) A(t)}_{0}}{\ex{A^\dagger(0) A(t)}_{0, \rm eq}} &=  \e^{\frac{\tau_s I_+}{e^*} {\rm Re}\mathcal{I}(t/\tau_s)} \Bigg\{
    \cos\left[\frac{\tau_s I_-}{e^*} {\rm Im}\mathcal{I}(t/\tau_s)\right]
    \notag \\&  \ \ \ \ \ \
    + i \sin\left[ \frac{\tau_s I_-}{e^*} {\rm Im}\mathcal{I}(t/\tau_s) \right] \Bigg\} \ , \label{Eq:AAexpSplit}
\end{align}
where $\ex{.}_0$ denotes the non-equilibrium expectation value in absence of the tunneling coupling,   and $\ex{A^\dagger(0)A(t)}_{0, \rm eq}=\tau_c^{2\delta} \e^{-i\pi\delta\,{\rm sgn}(t)}/|t|^{2\delta}$  is the equilibrium expectation value of the unperturbed edge with  short time cutoff $\tau_c$. We defined  $I_\pm=I_{u,0}\pm I_{d,0}$, where $I_{\alpha,0}$ is the current of the dilute beam on edge $\alpha$ before the QPC. 
In Eq.~(\ref{Eq:AAexpSplit}),  the integral $\mathcal{I}$ is given by
\begin{align}
    \mathcal{I}(t/\tau_s) &\equiv 
    \int_{-\infty}^{\infty} {\rm d}x \Bigg\{\exp\Bigg[i \frac{2\theta}{\pi}\Bigg( \arctan\left(x-\frac{t}{2\tau_s}\right)
     \notag \\& \ \ \ \ \ \
     -\arctan\left(x+\frac{t}{2\tau_s}\right) \Bigg)\Bigg] - 1\Bigg\} \ . 
     \label{Eq:Iintegral}
\end{align}
We used that the real part of the integral is an even function of $t$, while the imaginary part is odd, and  the contributions to the exponent of the correlation function for edge $\alpha$ is given by ${\rm Re}(\mathcal{I})+\alpha {\rm Im}(\mathcal{I})$ ($\alpha=\pm1$ for $u$ and  $d$ respectively).

By using the expression for the correlation functions Eq.~\eqref{Eq:AAexpSplit} in the current expectation value, we find to leading order in the  tunneling amplitude $\zeta$  
\begin{align}
    &\ex{\hat{I}_T} 
    =  -i e^* \frac{|\zeta|^2 \tau_c^{2\delta}}{\tau_s^{2\delta-1}} \int_{-\infty}^\infty {\rm d} \tilde{t} \frac{\sin\left(\frac{\tau_s}{e^*} I_- {\rm Im}\mathcal{I}(\tilde{t})\right)}{\exp\left(-\frac{\tau_s}{e^*} I_+ {\rm Re}\mathcal{I}(\tilde{t})\right)} 
    \notag \\& \ \times 
    \left[  \left(\frac{1}{i\tilde{t}+\tau_c/\tau_s}\right)^{2\delta}- \left(\frac{1}{-i\tilde{t}+\tau_c/\tau_s}\right)^{2\delta} \right].   \label{Eq:CurrentExp}
\end{align}
Similarly, we find that the current noise is given by
\begin{align}
    &\ex{(\delta\hat{I})^2}_{\omega=0}  
    = (e^*)^2  \frac{|\zeta|^2 \tau_c^{2\delta}}{\tau_s^{2\delta-1}} \int_{-\infty}^\infty {\rm d}\tilde{t}\,\frac{\cos\left(\frac{\tau_s}{e^*} I_- {\rm Im}\mathcal{I}(\tilde{t})\right)}{\exp\left(-\frac{\tau_s}{e^*} I_+ {\rm Re}\mathcal{I}(\tilde{t})\right)} 
    \notag \\&   \times 
    \left[ \left(\frac{1}{i \tilde{t} + \tau_c/\tau_s}\right)^{2\delta} + \left(\frac{1}{-i \tilde{t} + \tau_c/\tau_s}\right)^{2\delta}  \right].\!
\end{align}
In the exponential factors of the current and noise integrands, the transparency of the source QPC   
\begin{align}
    T_s &\equiv \frac{I_+}{e^*}\tau_s \ , 
    \label{Eq:Ts}
\end{align}
determines how fast the integrands decay.  Even  in the limit $T_s\to0$ where it seems plausible that tunneling onto edge $\alpha$ may be  described by a discrete step in the boson field, the short time behavior of the integrals is not accurately approximated. This problem arises for $\delta>1/2$, where the integrals have large short time contributions from the $(\pm i t+\tau_c/\tau_s)^{-2\delta}$ terms.  

We denote the ratio of short time cutoff and broadening time as $\eta=\tau_c/\tau_s$ and require $\eta<1$. We find below that already for  values of $\eta\sim 10^{-2}$-$10^{-1}$ the dependence of current and noise on $\eta$ is negligibly weak.

\emph{Fano factor---}As an experimentally accessible observable which depends  on the braiding phase $2\theta$, we consider the generalized Fano factor
\begin{align}
    P(I_-/I_+) &= \frac{\ex{\delta I_d \delta I_u}_{\omega=0}}{e^* (I_{u,0}\partial_{I_{u,0}} - I_{d,0}\partial_{I_{d,0}} ) \ex{\hat{I}_T}} \equiv\frac{P_n}{P_d} \ ,
\end{align}
where the denominator is the noise in incoming currents, transmitted through the QPC -- a generalization of Johnson-Nyquist noise. The fact that the denominator is defined as a derivative  of the current expectation value allows us to focus on the symmetric case, $I_-=0$, for which $\ex{\hat{I}_T}$ vanishes. It can be shown that the Fano factor only depends on the ratio $I_-/I_+$ and has a maximum at $I_-=0$ \cite{Rosenow.2016}. The numerator of the Fano factor contains the  sum of the 
current noise due to partitioning at the QPC and of the transmitted noise.  
The numerator  is given by $\ex{\delta I_d \delta I_u}_{\omega=0} = -\ex{\delta\hat{I}^2}+e^* (I_{u,0}\partial_{I_{u,0}} - I_{d,0}\partial_{I_{d,0}} ) \ex{\hat{I}_T}$ \cite{Rosenow.2016}. 

For the symmetric case $I_-=0$, the denominator of the Fano factor can be expressed as
\begin{align}
    &\frac{P_d(0)}{2e^*I_+|\zeta|^2 \tau_c^{2\delta}\tau_s^{-2(\delta-1)}}  
    = -  2\int_{0}^\infty {\rm d}\tilde{t}\, {\rm Im}[\mathcal{I}(\tilde{t})]
    \,\e^{T_s {\rm Re}[\mathcal{I}(\tilde{t})]} 
    \notag \\& \phantom{------} \times 
    \sin\left[2\delta\arctan\left(\frac{\tilde{t}}{\eta}\right)\right] \left[\tilde{t}^2+\eta^2\right]^{-\delta}  ,
\end{align}
and the difference between numerator and denominator, i.e., $-\ex{\delta \hat{I}^2}$, becomes
\begin{align}
    &\frac{P_n(0)-P_d(0)}{2e^* I_+  |\zeta|^2 \tau_c^{2\delta} \tau_s^{-(2\delta-2)}}   
    =  -\frac{2}{T_s}\int_{0}^\infty {\rm d}\tilde{t}\,\e^{T_s {\rm Re}[ \mathcal{I}(\tilde{t})]}  
    \notag \\&\phantom{------}  \times 
    \cos\left[2\delta\arctan\left(\frac{\tilde{t}}{\eta}\right)\right] \left[\tilde{t}^2+\eta^2\right]^{-\delta} \ .
\end{align}
These expressions can efficiently be numerically evaluated as they are real and the integral $\mathcal{I}$ only depend on a single parameter $\theta$ and the reduced time $\tilde{t}=t/\tau_s$, which allows  
reusing the results for different values of $\delta$ and $\eta$.

The full Fano factor in the symmetric case $I_-=0$ is then given by
\begin{widetext}
\begin{align}
    P(0) &=  1+\frac{1}{T_s} \frac{\int_{0}^\infty {\rm d}\tilde{t}\,\e^{T_s {\rm Re}[ \mathcal{I}(\tilde{t})]}   \cos\left[2\delta\arctan\left(\frac{\tilde{t}}{\eta}\right)\right] \left[\tilde{t}^2+\eta^2\right]^{-\delta}}{\int_{0}^\infty {\rm d}\tilde{t}\, {\rm Im}[\mathcal{I}(\tilde{t})]
    \,\e^{T_s {\rm Re}[\mathcal{I}(\tilde{t})]} 
    \, \sin\left[2\delta\arctan\left(\frac{\tilde{t}}{\eta}\right)\right] \left[\tilde{t}^2+\eta^2\right]^{-\delta}}\ .
\end{align}
\end{widetext}

\begin{figure}[t!]
    \centering
    \includegraphics[width=8.3cm]{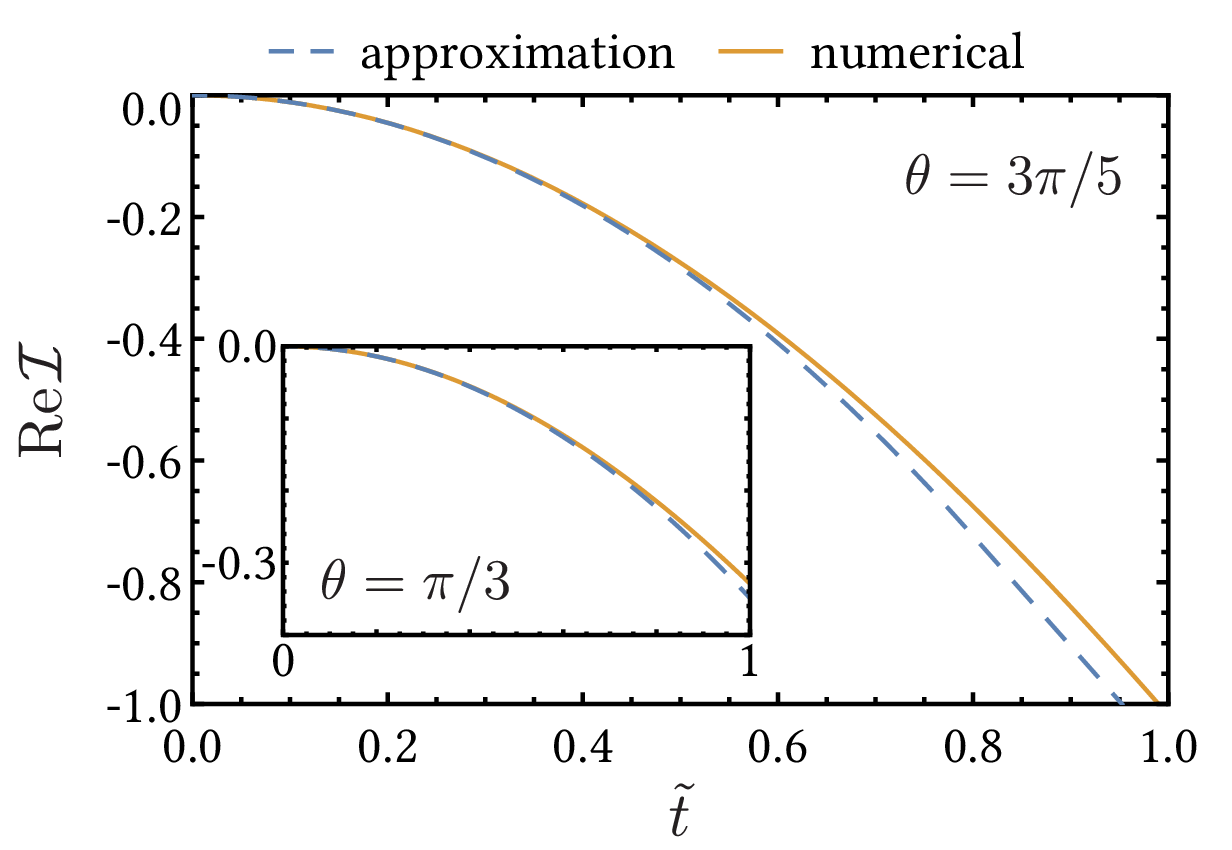}
    \caption{\label{Fig:ReIApprox}Analytic approximation (blue, dashed) of the real part ${\rm Re}\mathcal{I}(\tilde{t})$  at small times together with the numerically obtained real part of Eq.~\eqref{Eq:Iintegral} (orange). The main panel shows the comparison for the phase $\theta=3\pi/5$ and the inset for Laughlin anyons with $\theta=\pi/3$. We find that the approximation accurately describes the short time behavior. 
 }
\end{figure}

\emph{Analytic solution for $\delta=1$---}Before we discuss the numerical results for $P(0)$ as a function of $\delta$, we first present an approximate analytical solution for $\delta=1$, which is close to the experimentally estimated value of the dynamical exponent \cite{Lee.2023,Ruelle.2023}, and show that the Fano factor is non-universal and logarithmically diverging for $T_s\to0$ in this case. 

In the limit $\delta\to1$, the denominator of the Fano factor can be evaluated analytically as the expression in the square brackets of Eq.~\eqref{Eq:CurrentExp} reduces to the first derivative of a delta distribution $2\pi i \partial_{\tilde{t}}\delta(\tilde{t})$  and $\partial_{\tilde{t}} \mathcal{I}\mid_{\tilde{t}=0}   = -2i\theta$, such that
\begin{align} 
    \frac{ P_d(I_-=0;\delta=1)}{2e^*I_+|\zeta|^2\tau_c^{2}}&=2\pi \theta + \mathcal{O}(\tau_c)\ . \label{Eq:DenominatorDelta1}
\end{align}

To evaluate the numerator, we approximate the integral $\mathcal{I}$ for short times using a Taylor expansion to second order,  
    $ \mathcal{I}(\tilde{t}\to0) = -2i\theta \tilde{t} -  {\tilde{t}^2\theta^2}/{\pi} + \mathcal{O}(\tilde{t}^3)$,
and for late times using its asymptotic behavior
    $\mathcal{I}(\tilde{t}\to \infty) \sim  -\tilde{t} (\e^{2i\theta}-1) $.  
We smoothly connect the two approximations at the point $t_a = \pi(1-\cos(2\theta))/(2\theta^2)$, where the asymptotic slope agrees with the slope of the tangent of the quadratic behavior for small times, and obtain for the real part of the integral
\begin{align}
    {\rm Re}\mathcal{I}(\tilde{t} )&\approx \begin{cases}
        - {\tilde{t}^2\theta^2}/{\pi} & \text{ for } \tilde{t} < t_a \\
        -\tilde{t} [1-\cos(2\theta)] + t_a^2 \theta^2 /{\pi}
        & \text{ for } \tilde{t} \geq t_a  . \label{Eq:ReIApproximationFull}
    \end{cases}
\end{align}
In Fig.~\ref{Fig:ReIApprox}, we show the short time approximation together with the numerically obtained real part of $\mathcal{I}$ for both $\theta=\pi/3$ and $\theta=3\pi/5$. Using the approximation, we find the  Fano factor  
\begin{align}
    P(0;&\delta=1) \approx 1 + \frac{1-\cos2\theta}{\pi\theta} \bigg[1-\gamma 
    \notag \\& 
    - \ln\left(\frac{\pi(1-\cos2\theta)^2}{2\theta^2}\,T_s\right)\bigg] + \mathcal{O}(\eta, T_s) \ ,
\end{align} 
with the Euler constant $\gamma$. The logarithmic divergence for $T_s\to0$ indicates that the Fano factor for $\delta = 1$ depends in a significant way  on the source QPC transparency $T_s$.

\begin{figure}[t!]
    \centering
    \includegraphics[width=8.3cm]{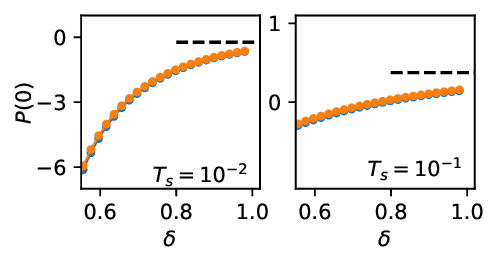}
    \caption{\label{Fig:Num35}Numerical results for the Fano factor $P(0)$ with symmetric currents $I_-=0$ for anyons with $\theta=3\pi/5$ as a function  of the dynamical exponent $\delta>1/2$. The orange dots show Fano factors for $\eta=\tau_c/\tau_s = 10^{-1}$, while the blue points correspond to $\eta=10^{-2}$. The dashed, black line shows the analytical approximation for $\delta=1$. In the left panel, we use finite width solitons with $T_s=10^{-2}$ and for the right panel $T_s=10^{-1}$.  The dependence on $\eta<1$ is negligible.  }
\end{figure}
\begin{figure}[t!]
    \centering
    \includegraphics[width=8.3cm]{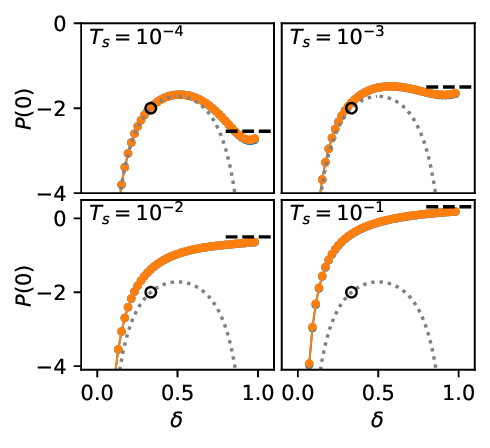}
    \caption{\label{Fig:Num13}Numerical results for the Fano factor $P(0)$ with symmetric currents $I_-=0$ for Laughlin anyons with $\theta=\pi/3$ as a function  of the dynamical exponent $\delta$. The orange dots show Fano factors for $\eta = 10^{-1}$, while the blue points correspond to $\eta=10^{-2}$. The dashed, black line shows the analytical approximation for $\delta=1$ and the dotted gray line depicts the Fano factor obtain for zero-width solitons in Ref.~\cite{Rosenow.2016}. For $\delta<1/2$ and $T_s\in\{10^{-4},10^{-3}\}$, we find that the finite soliton width  changes the Fano factor little as compared  to the zero-width prediction. For $\delta>1/2$, the Fano factor depends more strongly on the value of $T_s$, to a degree that the divergence towards $\delta=1$ is lifted by the finite soliton width. 
 }
\end{figure}

\emph{Numerical results for $\theta=3\pi/5$---}We  numerically compute the Fano factor for anyons with exchange phase $\theta=3\pi/5$.    As recent studies \cite{Ruelle.2023,Glidic.2023,Lee.2023} indicate that $\delta > 3/5$ may be realized experimentally, we compute the Fano factor for the symmetric case as a function of  $\delta$ (Fig.~\ref{Fig:Num35}). We find that for a dilution of $T_s = 10^{-2}$ (panel a), indeed a  negative Fano factor is obtained for all $\delta>1/2$, which  has a significant dependence on $\delta$. For even wider  solitons with $T_s=10^{-1}$, we find that the Fano factor changes sign as a function of $\delta$.  In Ref.~\cite{Ruelle.2023}, measurements have been performed for $T_s=0.14$ and $0.34$, in agreement with the theoretical finding of $P(0) = -0.21$ for 
$T_s = 0.1$ and $\delta = 3/5$. In addition, Ref.~\cite{Glidic.2023} reports results for dilutions in the range $T_s\in[0.25,0.5]$, somewhat outside the 
the dilute anyon regime discussed here.

\emph{Numerical results for $\theta=\pi/3$---}  For Laughlin anyons with $\theta=\pi/3$,  we expect to recover  the  result of a universal Fano factor from Ref.~\cite{Rosenow.2016} close to $\delta=1/3 < 1/2$. Indeed, we find that the results with finite soliton width show good agreement with the prediction for zero width for $\delta<1/2$ (Fig.~\ref{Fig:Num13}). For larger values of $\delta$ approaching one, the finite soliton width removes  the divergence to large negative values,  yielding negative Fano factors on the order of $-2$ for a dilution of $T_s = 10^{-3}$, compatible with recent experiments \cite{Bartolomei.2020,Lee.2023,Ruelle.2023,Glidic.2023}.

\emph{Discussion---}By considering anyon pulses with  finite dilution $10^{-2} \lesssim T_s \lesssim  10^{-1}$, we find a negative Fano factor in agreement with recent experimental results \cite{Glidic.2023,Ruelle.2023} for anyons with $e^* = e/5$  and $\theta=3\pi/5$. However, when further reducing $T_s$, unphysical jumps in the Fano factor 
as a function of $\delta$ appear for $\theta= 3 \pi/5$ (not shown in Fig.~\ref{Fig:Num35}), which are due to  a negative tunneling current that cannot be explained by numerical instabilities. In addition, while the $\theta=\pi/3$ results are in overall agreement with experiments, yielding an increasing Fano factor with increasing $T_s$, the dependence on $T_s$ seems to be somewhat  weaker in experiments \cite{Ruelle.2023} than predicted theoretically. 
 This could be due to the semi-classical description of non-equilibrium physics, or perhaps related to the  $\arctan$-shape of the solitons chosen in our calculation. 

\emph{Conclusion---}Our study reveals the crucial importance of finite soliton widths in analyzing anyon collider (time-domain interference)  experiments, especially for exchange phases greater than $\pi/2$ and dynamical exponents above 1/2. We demonstrate that incorporating a finite soliton width leads to a  prediction of  negative Fano factors for anyons with exchange phases of $3 \pi/5$, in agreement with recent experimental data. This finding is crucial for accurately interpreting  experiments and understanding universal and non-universal contributions to Fano factors in anyonic systems, thereby offering a more comprehensive framework for exploring anyonic statistics in fractional quantum Hall states.

\emph{Note added in proof---}After completion of this work, we became aware of Ref.~\cite{iyer2023finite}, which also studies the soliton shape for the description of anyon colliders and finds results consistent with ours.

\begin{acknowledgments}  
	\textit{Acknowledgments:}  We would thank X.-G.~Wen, N.~Schiller, G.~F\`eve, and F.~Pierre for helpful conversations.\\
\end{acknowledgments}


\begin{thebibliography}{31}%
    \makeatletter
    \providecommand \@ifxundefined [1]{%
     \@ifx{#1\undefined}
    }%
    \providecommand \@ifnum [1]{%
     \ifnum #1\expandafter \@firstoftwo
     \else \expandafter \@secondoftwo
     \fi
    }%
    \providecommand \@ifx [1]{%
     \ifx #1\expandafter \@firstoftwo
     \else \expandafter \@secondoftwo
     \fi
    }%
    \providecommand \natexlab [1]{#1}%
    \providecommand \enquote  [1]{``#1''}%
    \providecommand \bibnamefont  [1]{#1}%
    \providecommand \bibfnamefont [1]{#1}%
    \providecommand \citenamefont [1]{#1}%
    \providecommand \href@noop [0]{\@secondoftwo}%
    \providecommand \href [0]{\begingroup \@sanitize@url \@href}%
    \providecommand \@href[1]{\@@startlink{#1}\@@href}%
    \providecommand \@@href[1]{\endgroup#1\@@endlink}%
    \providecommand \@sanitize@url [0]{\catcode `\\12\catcode `\$12\catcode
      `\&12\catcode `\#12\catcode `\^12\catcode `\_12\catcode `\%12\relax}%
    \providecommand \@@startlink[1]{}%
    \providecommand \@@endlink[0]{}%
    \providecommand \url  [0]{\begingroup\@sanitize@url \@url }%
    \providecommand \@url [1]{\endgroup\@href {#1}{\urlprefix }}%
    \providecommand \urlprefix  [0]{URL }%
    \providecommand \Eprint [0]{\href }%
    \providecommand \doibase [0]{https://doi.org/}%
    \providecommand \selectlanguage [0]{\@gobble}%
    \providecommand \bibinfo  [0]{\@secondoftwo}%
    \providecommand \bibfield  [0]{\@secondoftwo}%
    \providecommand \translation [1]{[#1]}%
    \providecommand \BibitemOpen [0]{}%
    \providecommand \bibitemStop [0]{}%
    \providecommand \bibitemNoStop [0]{.\EOS\space}%
    \providecommand \EOS [0]{\spacefactor3000\relax}%
    \providecommand \BibitemShut  [1]{\csname bibitem#1\endcsname}%
    \let\auto@bib@innerbib\@empty
    \bibitem [{\citenamefont {Leinaas}\ and\ \citenamefont
      {Myrheim}(1977)}]{Leinaas.1977}%
      \BibitemOpen
      \bibfield  {author} {\bibinfo {author} {\bibfnamefont {J.}~\bibnamefont
      {Leinaas}}\ and\ \bibinfo {author} {\bibfnamefont {J.}~\bibnamefont
      {Myrheim}},\ }\href@noop {} {\bibfield  {journal} {\bibinfo  {journal} {Il
      nuovo cimento}\ }\textbf {\bibinfo {volume} {37}},\ \bibinfo {pages} {132}
      (\bibinfo {year} {1977})}\BibitemShut {NoStop}%
    \bibitem [{\citenamefont {Laughlin}(1983)}]{Laughlin.1983}%
      \BibitemOpen
      \bibfield  {author} {\bibinfo {author} {\bibfnamefont {R.~B.}\ \bibnamefont
      {Laughlin}},\ }\href@noop {} {\bibfield  {journal} {\bibinfo  {journal}
      {Physical Review Letters}\ }\textbf {\bibinfo {volume} {50}},\ \bibinfo
      {pages} {1395} (\bibinfo {year} {1983})}\BibitemShut {NoStop}%
    \bibitem [{\citenamefont {Halperin}(1984)}]{Halperin.1984}%
      \BibitemOpen
      \bibfield  {author} {\bibinfo {author} {\bibfnamefont {B.~I.}\ \bibnamefont
      {Halperin}},\ }\href@noop {} {\bibfield  {journal} {\bibinfo  {journal}
      {Physical Review Letters}\ }\textbf {\bibinfo {volume} {52}},\ \bibinfo
      {pages} {1583} (\bibinfo {year} {1984})}\BibitemShut {NoStop}%
    \bibitem [{\citenamefont {Arovas}\ \emph {et~al.}(1984)\citenamefont {Arovas},
      \citenamefont {Schrieffer},\ and\ \citenamefont {Wilczek}}]{Arovas.1984}%
      \BibitemOpen
      \bibfield  {author} {\bibinfo {author} {\bibfnamefont {D.}~\bibnamefont
      {Arovas}}, \bibinfo {author} {\bibfnamefont {J.~R.}\ \bibnamefont
      {Schrieffer}},\ and\ \bibinfo {author} {\bibfnamefont {F.}~\bibnamefont
      {Wilczek}},\ }\href@noop {} {\bibfield  {journal} {\bibinfo  {journal}
      {Physical review letters}\ }\textbf {\bibinfo {volume} {53}},\ \bibinfo
      {pages} {722} (\bibinfo {year} {1984})}\BibitemShut {NoStop}%
    \bibitem [{\citenamefont {Feldman}\ and\ \citenamefont
      {Halperin}(2021)}]{Feldman.2021}%
      \BibitemOpen
      \bibfield  {author} {\bibinfo {author} {\bibfnamefont {D.~E.}\ \bibnamefont
      {Feldman}}\ and\ \bibinfo {author} {\bibfnamefont {B.~I.}\ \bibnamefont
      {Halperin}},\ }\href@noop {} {\bibfield  {journal} {\bibinfo  {journal}
      {Reports on Progress in Physics}\ }\textbf {\bibinfo {volume} {84}},\
      \bibinfo {pages} {076501} (\bibinfo {year} {2021})}\BibitemShut {NoStop}%
    \bibitem [{\citenamefont {Reznikov}\ \emph {et~al.}(1999)\citenamefont
      {Reznikov}, \citenamefont {Picciotto}, \citenamefont {Griffiths},
      \citenamefont {Heiblum},\ and\ \citenamefont {Umansky}}]{Reznikov.1999}%
      \BibitemOpen
      \bibfield  {author} {\bibinfo {author} {\bibfnamefont {M.}~\bibnamefont
      {Reznikov}}, \bibinfo {author} {\bibfnamefont {R.~d.}\ \bibnamefont
      {Picciotto}}, \bibinfo {author} {\bibfnamefont {T.}~\bibnamefont
      {Griffiths}}, \bibinfo {author} {\bibfnamefont {M.}~\bibnamefont {Heiblum}},\
      and\ \bibinfo {author} {\bibfnamefont {V.}~\bibnamefont {Umansky}},\
      }\href@noop {} {\bibfield  {journal} {\bibinfo  {journal} {Nature}\ }\textbf
      {\bibinfo {volume} {399}},\ \bibinfo {pages} {238} (\bibinfo {year}
      {1999})}\BibitemShut {NoStop}%
    \bibitem [{\citenamefont {Saminadayar}\ \emph {et~al.}(1997)\citenamefont
      {Saminadayar}, \citenamefont {Glattli}, \citenamefont {Jin},\ and\
      \citenamefont {Etienne}}]{Saminadayar.1997}%
      \BibitemOpen
      \bibfield  {author} {\bibinfo {author} {\bibfnamefont {L.}~\bibnamefont
      {Saminadayar}}, \bibinfo {author} {\bibfnamefont {D.}~\bibnamefont
      {Glattli}}, \bibinfo {author} {\bibfnamefont {Y.}~\bibnamefont {Jin}},\ and\
      \bibinfo {author} {\bibfnamefont {B.~c.-m.}\ \bibnamefont {Etienne}},\
      }\href@noop {} {\bibfield  {journal} {\bibinfo  {journal} {Physical Review
      Letters}\ }\textbf {\bibinfo {volume} {79}},\ \bibinfo {pages} {2526}
      (\bibinfo {year} {1997})}\BibitemShut {NoStop}%
    \bibitem [{\citenamefont {Bartolomei}\ \emph {et~al.}(2020)\citenamefont
      {Bartolomei}, \citenamefont {Kumar}, \citenamefont {Bisognin}, \citenamefont
      {Marguerite}, \citenamefont {Berroir}, \citenamefont {Bocquillon},
      \citenamefont {Placais}, \citenamefont {Cavanna}, \citenamefont {Dong},
      \citenamefont {Gennser} \emph {et~al.}}]{Bartolomei.2020}%
      \BibitemOpen
      \bibfield  {author} {\bibinfo {author} {\bibfnamefont {H.}~\bibnamefont
      {Bartolomei}}, \bibinfo {author} {\bibfnamefont {M.}~\bibnamefont {Kumar}},
      \bibinfo {author} {\bibfnamefont {R.}~\bibnamefont {Bisognin}}, \bibinfo
      {author} {\bibfnamefont {A.}~\bibnamefont {Marguerite}}, \bibinfo {author}
      {\bibfnamefont {J.-M.}\ \bibnamefont {Berroir}}, \bibinfo {author}
      {\bibfnamefont {E.}~\bibnamefont {Bocquillon}}, \bibinfo {author}
      {\bibfnamefont {B.}~\bibnamefont {Placais}}, \bibinfo {author} {\bibfnamefont
      {A.}~\bibnamefont {Cavanna}}, \bibinfo {author} {\bibfnamefont
      {Q.}~\bibnamefont {Dong}}, \bibinfo {author} {\bibfnamefont {U.}~\bibnamefont
      {Gennser}}, \emph {et~al.},\ }\href@noop {} {\bibfield  {journal} {\bibinfo
      {journal} {Science}\ }\textbf {\bibinfo {volume} {368}},\ \bibinfo {pages}
      {173} (\bibinfo {year} {2020})}\BibitemShut {NoStop}%
    \bibitem [{\citenamefont {Nakamura}\ \emph {et~al.}(2020)\citenamefont
      {Nakamura}, \citenamefont {Liang}, \citenamefont {Gardner},\ and\
      \citenamefont {Manfra}}]{Nakamura.2020}%
      \BibitemOpen
      \bibfield  {author} {\bibinfo {author} {\bibfnamefont {J.}~\bibnamefont
      {Nakamura}}, \bibinfo {author} {\bibfnamefont {S.}~\bibnamefont {Liang}},
      \bibinfo {author} {\bibfnamefont {G.~C.}\ \bibnamefont {Gardner}},\ and\
      \bibinfo {author} {\bibfnamefont {M.~J.}\ \bibnamefont {Manfra}},\
      }\href@noop {} {\bibfield  {journal} {\bibinfo  {journal} {Nature Physics}\
      }\textbf {\bibinfo {volume} {16}},\ \bibinfo {pages} {931} (\bibinfo {year}
      {2020})}\BibitemShut {NoStop}%
    \bibitem [{\citenamefont {Nakamura}\ \emph {et~al.}(2022)\citenamefont
      {Nakamura}, \citenamefont {Liang}, \citenamefont {Gardner},\ and\
      \citenamefont {Manfra}}]{Nakamura.2022}%
      \BibitemOpen
      \bibfield  {author} {\bibinfo {author} {\bibfnamefont {J.}~\bibnamefont
      {Nakamura}}, \bibinfo {author} {\bibfnamefont {S.}~\bibnamefont {Liang}},
      \bibinfo {author} {\bibfnamefont {G.~C.}\ \bibnamefont {Gardner}},\ and\
      \bibinfo {author} {\bibfnamefont {M.~J.}\ \bibnamefont {Manfra}},\
      }\href@noop {} {\bibfield  {journal} {\bibinfo  {journal} {Nature
      Communications}\ }\textbf {\bibinfo {volume} {13}},\ \bibinfo {pages} {344}
      (\bibinfo {year} {2022})}\BibitemShut {NoStop}%
    \bibitem [{\citenamefont {Nakamura}\ \emph {et~al.}(2023)\citenamefont
      {Nakamura}, \citenamefont {Liang}, \citenamefont {Gardner},\ and\
      \citenamefont {Manfra}}]{Nakamura.2023}%
      \BibitemOpen
      \bibfield  {author} {\bibinfo {author} {\bibfnamefont {J.}~\bibnamefont
      {Nakamura}}, \bibinfo {author} {\bibfnamefont {S.}~\bibnamefont {Liang}},
      \bibinfo {author} {\bibfnamefont {G.~C.}\ \bibnamefont {Gardner}},\ and\
      \bibinfo {author} {\bibfnamefont {M.~J.}\ \bibnamefont {Manfra}},\ }\href
      {https://doi.org/10.1103/PhysRevX.13.041012} {\bibfield  {journal} {\bibinfo
      {journal} {Phys. Rev. X}\ }\textbf {\bibinfo {volume} {13}},\ \bibinfo
      {pages} {041012} (\bibinfo {year} {2023})}\BibitemShut {NoStop}%
    \bibitem [{\citenamefont {Lee}\ \emph {et~al.}(2023)\citenamefont {Lee},
      \citenamefont {Hong}, \citenamefont {Alkalay}, \citenamefont {Schiller},
      \citenamefont {Umansky}, \citenamefont {Heiblum}, \citenamefont {Oreg},\ and\
      \citenamefont {Sim}}]{Lee.2023}%
      \BibitemOpen
      \bibfield  {author} {\bibinfo {author} {\bibfnamefont {J.-Y.~M.}\
      \bibnamefont {Lee}}, \bibinfo {author} {\bibfnamefont {C.}~\bibnamefont
      {Hong}}, \bibinfo {author} {\bibfnamefont {T.}~\bibnamefont {Alkalay}},
      \bibinfo {author} {\bibfnamefont {N.}~\bibnamefont {Schiller}}, \bibinfo
      {author} {\bibfnamefont {V.}~\bibnamefont {Umansky}}, \bibinfo {author}
      {\bibfnamefont {M.}~\bibnamefont {Heiblum}}, \bibinfo {author} {\bibfnamefont
      {Y.}~\bibnamefont {Oreg}},\ and\ \bibinfo {author} {\bibfnamefont {H.-S.}\
      \bibnamefont {Sim}},\ }\href@noop {} {\bibfield  {journal} {\bibinfo
      {journal} {Nature}\ }\textbf {\bibinfo {volume} {617}},\ \bibinfo {pages}
      {277} (\bibinfo {year} {2023})}\BibitemShut {NoStop}%
    \bibitem [{\citenamefont {Ruelle}\ \emph {et~al.}(2023)\citenamefont {Ruelle},
      \citenamefont {Frigerio}, \citenamefont {Berroir}, \citenamefont
      {Pla{\c{c}}ais}, \citenamefont {Rech}, \citenamefont {Cavanna}, \citenamefont
      {Gennser}, \citenamefont {Jin},\ and\ \citenamefont
      {F{\`e}ve}}]{Ruelle.2023}%
      \BibitemOpen
      \bibfield  {author} {\bibinfo {author} {\bibfnamefont {M.}~\bibnamefont
      {Ruelle}}, \bibinfo {author} {\bibfnamefont {E.}~\bibnamefont {Frigerio}},
      \bibinfo {author} {\bibfnamefont {J.-M.}\ \bibnamefont {Berroir}}, \bibinfo
      {author} {\bibfnamefont {B.}~\bibnamefont {Pla{\c{c}}ais}}, \bibinfo {author}
      {\bibfnamefont {J.}~\bibnamefont {Rech}}, \bibinfo {author} {\bibfnamefont
      {A.}~\bibnamefont {Cavanna}}, \bibinfo {author} {\bibfnamefont
      {U.}~\bibnamefont {Gennser}}, \bibinfo {author} {\bibfnamefont
      {Y.}~\bibnamefont {Jin}},\ and\ \bibinfo {author} {\bibfnamefont
      {G.}~\bibnamefont {F{\`e}ve}},\ }\href@noop {} {\bibfield  {journal}
      {\bibinfo  {journal} {Physical Review X}\ }\textbf {\bibinfo {volume} {13}},\
      \bibinfo {pages} {011031} (\bibinfo {year} {2023})}\BibitemShut {NoStop}%
    \bibitem [{\citenamefont {Glidic}\ \emph {et~al.}(2023)\citenamefont {Glidic},
      \citenamefont {Maillet}, \citenamefont {Aassime}, \citenamefont {Piquard},
      \citenamefont {Cavanna}, \citenamefont {Gennser}, \citenamefont {Jin},
      \citenamefont {Anthore},\ and\ \citenamefont {Pierre}}]{Glidic.2023}%
      \BibitemOpen
      \bibfield  {author} {\bibinfo {author} {\bibfnamefont {P.}~\bibnamefont
      {Glidic}}, \bibinfo {author} {\bibfnamefont {O.}~\bibnamefont {Maillet}},
      \bibinfo {author} {\bibfnamefont {A.}~\bibnamefont {Aassime}}, \bibinfo
      {author} {\bibfnamefont {C.}~\bibnamefont {Piquard}}, \bibinfo {author}
      {\bibfnamefont {A.}~\bibnamefont {Cavanna}}, \bibinfo {author} {\bibfnamefont
      {U.}~\bibnamefont {Gennser}}, \bibinfo {author} {\bibfnamefont
      {Y.}~\bibnamefont {Jin}}, \bibinfo {author} {\bibfnamefont {A.}~\bibnamefont
      {Anthore}},\ and\ \bibinfo {author} {\bibfnamefont {F.}~\bibnamefont
      {Pierre}},\ }\href@noop {} {\bibfield  {journal} {\bibinfo  {journal}
      {Physical Review X}\ }\textbf {\bibinfo {volume} {13}},\ \bibinfo {pages}
      {011030} (\bibinfo {year} {2023})}\BibitemShut {NoStop}%
    \bibitem [{\citenamefont {Safi}\ \emph {et~al.}(2001)\citenamefont {Safi},
      \citenamefont {Devillard},\ and\ \citenamefont {Martin}}]{Safi.2001}%
      \BibitemOpen
      \bibfield  {author} {\bibinfo {author} {\bibfnamefont {I.}~\bibnamefont
      {Safi}}, \bibinfo {author} {\bibfnamefont {P.}~\bibnamefont {Devillard}},\
      and\ \bibinfo {author} {\bibfnamefont {T.}~\bibnamefont {Martin}},\
      }\href@noop {} {\bibfield  {journal} {\bibinfo  {journal} {Physical Review
      Letters}\ }\textbf {\bibinfo {volume} {86}},\ \bibinfo {pages} {4628}
      (\bibinfo {year} {2001})}\BibitemShut {NoStop}%
    \bibitem [{\citenamefont {Vishveshwara}(2003)}]{Vishveshwara.2003}%
      \BibitemOpen
      \bibfield  {author} {\bibinfo {author} {\bibfnamefont {S.}~\bibnamefont
      {Vishveshwara}},\ }\href@noop {} {\bibfield  {journal} {\bibinfo  {journal}
      {Physical review letters}\ }\textbf {\bibinfo {volume} {91}},\ \bibinfo
      {pages} {196803} (\bibinfo {year} {2003})}\BibitemShut {NoStop}%
    \bibitem [{\citenamefont {Martin}(2005)}]{Martin.2005}%
      \BibitemOpen
      \bibfield  {author} {\bibinfo {author} {\bibfnamefont {T.}~\bibnamefont
      {Martin}},\ }\href@noop {} {\bibinfo {title} {{Noise in mesoscopic physics
      Nanophysics: Coherence and Transport ed H Bouchiat, Y Gefen, S Gu{\'e}ron, G
      Montambaux and J Dalibard}}} (\bibinfo {year} {2005})\BibitemShut {NoStop}%
    \bibitem [{\citenamefont {Kim}\ \emph {et~al.}(2005)\citenamefont {Kim},
      \citenamefont {Lawler}, \citenamefont {Vishveshwara},\ and\ \citenamefont
      {Fradkin}}]{Kim.2005}%
      \BibitemOpen
      \bibfield  {author} {\bibinfo {author} {\bibfnamefont {E.-A.}\ \bibnamefont
      {Kim}}, \bibinfo {author} {\bibfnamefont {M.}~\bibnamefont {Lawler}},
      \bibinfo {author} {\bibfnamefont {S.}~\bibnamefont {Vishveshwara}},\ and\
      \bibinfo {author} {\bibfnamefont {E.}~\bibnamefont {Fradkin}},\ }\href@noop
      {} {\bibfield  {journal} {\bibinfo  {journal} {Physical review letters}\
      }\textbf {\bibinfo {volume} {95}},\ \bibinfo {pages} {176402} (\bibinfo
      {year} {2005})}\BibitemShut {NoStop}%
    \bibitem [{\citenamefont {Vishveshwara}\ and\ \citenamefont
      {Cooper}(2010)}]{Vishveshwara.2010}%
      \BibitemOpen
      \bibfield  {author} {\bibinfo {author} {\bibfnamefont {S.}~\bibnamefont
      {Vishveshwara}}\ and\ \bibinfo {author} {\bibfnamefont {N.}~\bibnamefont
      {Cooper}},\ }\href@noop {} {\bibfield  {journal} {\bibinfo  {journal}
      {Physical Review B}\ }\textbf {\bibinfo {volume} {81}},\ \bibinfo {pages}
      {201306} (\bibinfo {year} {2010})}\BibitemShut {NoStop}%
    \bibitem [{\citenamefont {Campagnano}\ \emph {et~al.}(2012)\citenamefont
      {Campagnano}, \citenamefont {Zilberberg}, \citenamefont {Gornyi},
      \citenamefont {Feldman}, \citenamefont {Potter},\ and\ \citenamefont
      {Gefen}}]{Campagnano.2012}%
      \BibitemOpen
      \bibfield  {author} {\bibinfo {author} {\bibfnamefont {G.}~\bibnamefont
      {Campagnano}}, \bibinfo {author} {\bibfnamefont {O.}~\bibnamefont
      {Zilberberg}}, \bibinfo {author} {\bibfnamefont {I.~V.}\ \bibnamefont
      {Gornyi}}, \bibinfo {author} {\bibfnamefont {D.~E.}\ \bibnamefont {Feldman}},
      \bibinfo {author} {\bibfnamefont {A.~C.}\ \bibnamefont {Potter}},\ and\
      \bibinfo {author} {\bibfnamefont {Y.}~\bibnamefont {Gefen}},\ }\href@noop {}
      {\bibfield  {journal} {\bibinfo  {journal} {Physical review letters}\
      }\textbf {\bibinfo {volume} {109}},\ \bibinfo {pages} {106802} (\bibinfo
      {year} {2012})}\BibitemShut {NoStop}%
    \bibitem [{\citenamefont {Campagnano}\ \emph {et~al.}(2013)\citenamefont
      {Campagnano}, \citenamefont {Zilberberg}, \citenamefont {Gornyi},\ and\
      \citenamefont {Gefen}}]{Campagnano.2013}%
      \BibitemOpen
      \bibfield  {author} {\bibinfo {author} {\bibfnamefont {G.}~\bibnamefont
      {Campagnano}}, \bibinfo {author} {\bibfnamefont {O.}~\bibnamefont
      {Zilberberg}}, \bibinfo {author} {\bibfnamefont {I.~V.}\ \bibnamefont
      {Gornyi}},\ and\ \bibinfo {author} {\bibfnamefont {Y.}~\bibnamefont
      {Gefen}},\ }\href@noop {} {\bibfield  {journal} {\bibinfo  {journal}
      {Physical Review B}\ }\textbf {\bibinfo {volume} {88}},\ \bibinfo {pages}
      {235415} (\bibinfo {year} {2013})}\BibitemShut {NoStop}%
    \bibitem [{\citenamefont {Rosenow}\ \emph {et~al.}(2016)\citenamefont
      {Rosenow}, \citenamefont {Levkivskyi},\ and\ \citenamefont
      {Halperin}}]{Rosenow.2016}%
      \BibitemOpen
      \bibfield  {author} {\bibinfo {author} {\bibfnamefont {B.}~\bibnamefont
      {Rosenow}}, \bibinfo {author} {\bibfnamefont {I.~P.}\ \bibnamefont
      {Levkivskyi}},\ and\ \bibinfo {author} {\bibfnamefont {B.~I.}\ \bibnamefont
      {Halperin}},\ }\href@noop {} {\bibfield  {journal} {\bibinfo  {journal}
      {Physical Review Letters}\ }\textbf {\bibinfo {volume} {116}},\ \bibinfo
      {pages} {156802} (\bibinfo {year} {2016})}\BibitemShut {NoStop}%
    \bibitem [{\citenamefont {Kim}\ \emph {et~al.}(2006)\citenamefont {Kim},
      \citenamefont {Lawler}, \citenamefont {Vishveshwara},\ and\ \citenamefont
      {Fradkin}}]{Kim.2006}%
      \BibitemOpen
      \bibfield  {author} {\bibinfo {author} {\bibfnamefont {E.-A.}\ \bibnamefont
      {Kim}}, \bibinfo {author} {\bibfnamefont {M.~J.}\ \bibnamefont {Lawler}},
      \bibinfo {author} {\bibfnamefont {S.}~\bibnamefont {Vishveshwara}},\ and\
      \bibinfo {author} {\bibfnamefont {E.}~\bibnamefont {Fradkin}},\ }\href@noop
      {} {\bibfield  {journal} {\bibinfo  {journal} {Physical Review B}\ }\textbf
      {\bibinfo {volume} {74}},\ \bibinfo {pages} {155324} (\bibinfo {year}
      {2006})}\BibitemShut {NoStop}%
    \bibitem [{\citenamefont {Han}\ \emph {et~al.}(2016)\citenamefont {Han},
      \citenamefont {Park}, \citenamefont {Gefen},\ and\ \citenamefont
      {Sim}}]{Han.2016}%
      \BibitemOpen
      \bibfield  {author} {\bibinfo {author} {\bibfnamefont {C.}~\bibnamefont
      {Han}}, \bibinfo {author} {\bibfnamefont {J.}~\bibnamefont {Park}}, \bibinfo
      {author} {\bibfnamefont {Y.}~\bibnamefont {Gefen}},\ and\ \bibinfo {author}
      {\bibfnamefont {H.-S.}\ \bibnamefont {Sim}},\ }\href@noop {} {\bibfield
      {journal} {\bibinfo  {journal} {Nature communications}\ }\textbf {\bibinfo
      {volume} {7}},\ \bibinfo {pages} {11131} (\bibinfo {year}
      {2016})}\BibitemShut {NoStop}%
    \bibitem [{\citenamefont {Lee}\ \emph {et~al.}(2019)\citenamefont {Lee},
      \citenamefont {Han},\ and\ \citenamefont {Sim}}]{Lee.2019}%
      \BibitemOpen
      \bibfield  {author} {\bibinfo {author} {\bibfnamefont {B.}~\bibnamefont
      {Lee}}, \bibinfo {author} {\bibfnamefont {C.}~\bibnamefont {Han}},\ and\
      \bibinfo {author} {\bibfnamefont {H.-S.}\ \bibnamefont {Sim}},\ }\href@noop
      {} {\bibfield  {journal} {\bibinfo  {journal} {Physical Review Letters}\
      }\textbf {\bibinfo {volume} {123}},\ \bibinfo {pages} {016803} (\bibinfo
      {year} {2019})}\BibitemShut {NoStop}%
    \bibitem [{\citenamefont {Lee}\ \emph {et~al.}(2020)\citenamefont {Lee},
      \citenamefont {Han},\ and\ \citenamefont {Sim}}]{Lee.2020}%
      \BibitemOpen
      \bibfield  {author} {\bibinfo {author} {\bibfnamefont {J.-Y.~M.}\
      \bibnamefont {Lee}}, \bibinfo {author} {\bibfnamefont {C.}~\bibnamefont
      {Han}},\ and\ \bibinfo {author} {\bibfnamefont {H.-S.}\ \bibnamefont {Sim}},\
      }\href@noop {} {\bibfield  {journal} {\bibinfo  {journal} {Physical Review
      Letters}\ }\textbf {\bibinfo {volume} {125}},\ \bibinfo {pages} {196802}
      (\bibinfo {year} {2020})}\BibitemShut {NoStop}%
    \bibitem [{\citenamefont {Schiller}\ \emph {et~al.}(2023)\citenamefont
      {Schiller}, \citenamefont {Shapira}, \citenamefont {Stern},\ and\
      \citenamefont {Oreg}}]{Schiller.2022}%
      \BibitemOpen
      \bibfield  {author} {\bibinfo {author} {\bibfnamefont {N.}~\bibnamefont
      {Schiller}}, \bibinfo {author} {\bibfnamefont {Y.}~\bibnamefont {Shapira}},
      \bibinfo {author} {\bibfnamefont {A.}~\bibnamefont {Stern}},\ and\ \bibinfo
      {author} {\bibfnamefont {Y.}~\bibnamefont {Oreg}},\ }\href@noop {} {\bibfield
       {journal} {\bibinfo  {journal} {Physical Review Letters}\ }\textbf {\bibinfo
      {volume} {131}},\ \bibinfo {pages} {186601} (\bibinfo {year}
      {2023})}\BibitemShut {NoStop}%
    \bibitem [{\citenamefont {Halperin}(1982)}]{Halperin.1982}%
      \BibitemOpen
      \bibfield  {author} {\bibinfo {author} {\bibfnamefont {B.~I.}\ \bibnamefont
      {Halperin}},\ }\href@noop {} {\bibfield  {journal} {\bibinfo  {journal}
      {Physical review B}\ }\textbf {\bibinfo {volume} {25}},\ \bibinfo {pages}
      {2185} (\bibinfo {year} {1982})}\BibitemShut {NoStop}%
    \bibitem [{\citenamefont {Wen}(1990)}]{Wen.1990}%
      \BibitemOpen
      \bibfield  {author} {\bibinfo {author} {\bibfnamefont {X.-G.}\ \bibnamefont
      {Wen}},\ }\href@noop {} {\bibfield  {journal} {\bibinfo  {journal} {Physical
      Review B}\ }\textbf {\bibinfo {volume} {41}},\ \bibinfo {pages} {12838}
      (\bibinfo {year} {1990})}\BibitemShut {NoStop}%
    \bibitem [{\citenamefont {Rosenow}\ and\ \citenamefont
      {Halperin}(2002)}]{Rosenow.2002}%
      \BibitemOpen
      \bibfield  {author} {\bibinfo {author} {\bibfnamefont {B.}~\bibnamefont
      {Rosenow}}\ and\ \bibinfo {author} {\bibfnamefont {B.~I.}\ \bibnamefont
      {Halperin}},\ }\href {https://doi.org/10.1103/PhysRevLett.88.096404}
      {\bibfield  {journal} {\bibinfo  {journal} {Phys. Rev. Lett.}\ }\textbf
      {\bibinfo {volume} {88}},\ \bibinfo {pages} {096404} (\bibinfo {year}
      {2002})}\BibitemShut {NoStop}%
    \bibitem [{\citenamefont {Iyer}\ \emph {et~al.}(2023)\citenamefont {Iyer},
      \citenamefont {Ronetti}, \citenamefont {Gr{\'e}maud}, \citenamefont {Martin},
      \citenamefont {Rech},\ and\ \citenamefont {Jonckheere}}]{iyer2023finite}%
      \BibitemOpen
      \bibfield  {author} {\bibinfo {author} {\bibfnamefont {K.}~\bibnamefont
      {Iyer}}, \bibinfo {author} {\bibfnamefont {F.}~\bibnamefont {Ronetti}},
      \bibinfo {author} {\bibfnamefont {B.}~\bibnamefont {Gr{\'e}maud}}, \bibinfo
      {author} {\bibfnamefont {T.}~\bibnamefont {Martin}}, \bibinfo {author}
      {\bibfnamefont {J.}~\bibnamefont {Rech}},\ and\ \bibinfo {author}
      {\bibfnamefont {T.}~\bibnamefont {Jonckheere}},\ }\href@noop {} {\bibfield
      {journal} {\bibinfo  {journal} {arXiv preprint arXiv:2311.15094}\ } (\bibinfo
      {year} {2023})}\BibitemShut {NoStop}%
    \end{thebibliography}
%

\end{document}